\documentclass[12pt]{article}
\usepackage{epsfig,graphics,graphicx,graphpap,color}
\usepackage{setspace}
\newcommand\bmat{\left( \begin{array}{cc}}
\newcommand\emat{\end{array}\right)}

\def\msbar{\ifmmode{\overline{\rm MS}} \else{$\overline{\rm MS}$} \fi}
\def\drbar{\ifmmode{\overline{\rm DR}} \else{$\overline{\rm DR}$} \fi}
\def\beq      {\begin{equation}}
\def\eeq      {\end{equation}}
\def\ti              {\tilde}

\def\x               {\chi}

\def\ch                {{\ti \chi}}
\newcommand{\cha}[1]   {{\ti \x^+_{#1}}}

\newcommand{\neu}[1]   {{\ti \x^0_{#1}}}

\def\gev             {{\rm GeV}}

\def\drbarp{\ifmmode{\overline{\rm DR'}} \else{$\overline{\rm DR}'$} \fi}

\begin{document}
\pagestyle{empty} \vspace*{-1cm}
\begin{flushright}
  HEPHY-PUB 839/07 \\
\end{flushright}
\vspace*{2cm}
\begin{center}
{\Large\bf\boldmath
   Two-loop SUSY QCD corrections to the chargino masses in the MSSM}
   \\[5mm]
\vspace{10mm}
$\mbox{R. Sch\"ofbeck  and H. Eberl}$\\[5mm]
\vspace{6mm} $$ \mbox{{\it Institut f\"ur
Hochenergiephysik der \"Osterreichischen Akademie der
Wissenschaften,}}$$\vspace{-0.9cm} $$\mbox{{\it A--1050 Vienna,
Austria}}$$
\end{center}

\vspace{20mm}
\begin{abstract}
We have calculated the two-loop strong interaction corrections to the chargino pole masses in the \drbarp scheme in the
Minimal Supersymmetric Standard Model (MSSM) with complex parameters.
We have performed a detailed numerical analysis for a particular point in the parameter space
and found corrections of a few tenths of a percent.
We provide a computer program which calculates chargino and neutralino masses with complex parameters up to $\mathcal O(\alpha\alpha_S)$.

\end{abstract}
\vfill
\newpage
\pagestyle{plain} \setcounter{page}{2}

\section{Introduction}

 In the Minimal Supersymmetric Standard Model (MSSM), there are two charginos $\ch_1^\pm$ and $\ch_2^\pm$, which
are the fermion mass eigenstates of the supersymmetric partners of the $W^\pm$ and the charged Higgs bosons $H^\pm_{1,2}$.
Likewise, there are four neutralinos $\neu 1,\ldots,\neu 4$, which are the fermion mass eigenstates of the supersymmetric partners
of the photon, the $Z^0$-boson, and the neutral Higgs bosons $H^0_{1,2}$. Their mass matrices stem from soft gaugino breaking terms,
spontaneous symmetry breaking in the Higgs sector and in case of $\mu$ from the super-potential.

The next generation of future high-energy physics experiments at Tevatron, LHC and a future
$e^+e^-$ linear collider (ILC) will hopefully discover these particles if supersymmetry (SUSY) is realized at low energies.
Much work has been devoted  to the study of the physics interplay of experiments at LHC and ILC \cite{Weiglein:2004hn}.
Particularly at a linear collider, it will be possible to perform
measurements with high precision \cite{ Weiglein:2004hn, tesla, lincol}.
In fact, the accuracies
of the masses of the lighter SUSY fermions are in the permille region which makes
the inclusion of higher order corrections indispensible.

In the framework of the real MSSM important results on quark self-energies were obtained in \cite{Bednyakov:2002sf}-\cite{Bednyakov:2005kt}.
In \cite{Martin:2005ch,Yamada:2006vn} the gluino pole mass was calculated to two-loop order. Moreover, the MSSM Higgs-sector has
been studied in detail, even in the full complex model \cite{Heinemeyer:2007gj}-\cite{Heinemeyer:1998yj}.

In a previous work \cite{Schofbeck:2006gs} we studied loop corrections to the neutralino pole masses and found that
the relation between the $\drbar$-input and the physical observables has to be established at least at the two-loop level in order to match
experimental precision. Following these lines, we
calculate in this paper the two-loop $\mathcal O(\alpha\alpha_S)$ corrections to the charginos within the MSSM. We conclude that these two-loop corrections
are in the magnitude of the experimental uncertainty and therefore they are relevant e.g. for global fits of $\drbar$-parameters.

A new feature of this work is the inclusion of complex parameters in the MSSM. We therefore not only study the charginos but
also re-analyze the neutralino-masses with complex parameters and study in particular the dependence on the phase
of the soft trilinear breaking parameter $A_t$.

Generic analytic formulae for SUSY QCD corrections in $\mathcal O(\alpha\alpha_S)$ to fermion pole masses in the MSSM were already derived
in \cite{Martin:2005ch}. Our calculation is, however, completely independent. More precisely, we use semi-automatic {\sc Mathematica}
tools \cite{feynarts, feyncalc, tarcer} for the diagram generation and analytic simplifications.

In the Appendix we briefly describe our C-program {\sc Polxino} \cite{Polxino} developed for the calculation of chargino and neutralino
masses with complex parameters up to $\mathcal O(\alpha\alpha_S)$ with a convenient
{\sc SLHA}-interface \cite{Skands:2003cj} for numerical studies.

\section{Diagrammatics}

\begin{figure}[p]
\begin{center}
\begin{picture}(125,50)(0,0)
    \put(-19,-8){\mbox{\resizebox{!}{5cm}{\includegraphics{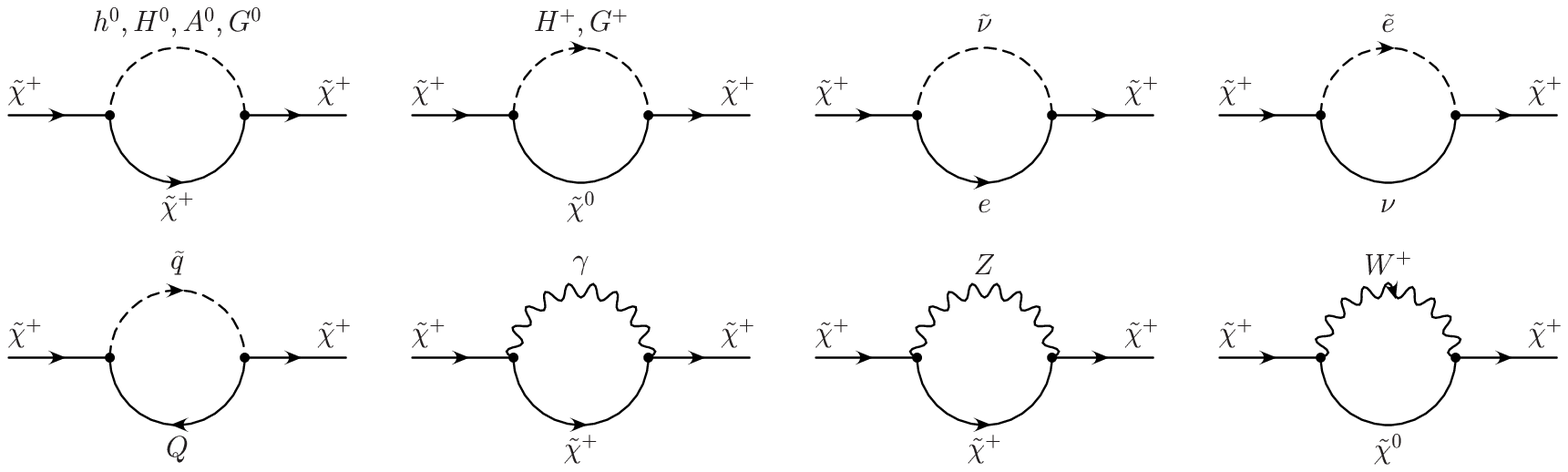}}}}
\end{picture}
\end{center}
\caption{\it Chargino one-loop self-energy diagrams}\label{Ch1LoopDiags}
\end{figure}

In Fig.~\ref{Ch1LoopDiags} we show all one-loop diagrams.
Similar to the neutralino case we checked our analytic one-loop calculation against previous work
\cite{Oller:2003ge, Fritzsche:2004ek} in the on-shell scheme and found agreement.

\begin{figure}[t]
\begin{center}
\begin{picture}(125,30)(0,0)
   \put(20,-10){\mbox{\resizebox{!}{3.3cm}{\includegraphics{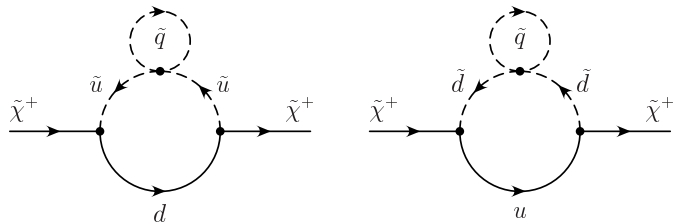}}}}
\end{picture}
\end{center}
\caption{\it  Chargino two-loop selfenergy diagrams with three inner squark lines}\label{Ch2LSquark}
\end{figure}

\begin{figure}[t]
\begin{center}
\begin{picture}(125,25)(0,0)
    \put(-19,-8){\mbox{\resizebox{!}{2.95cm}{\includegraphics{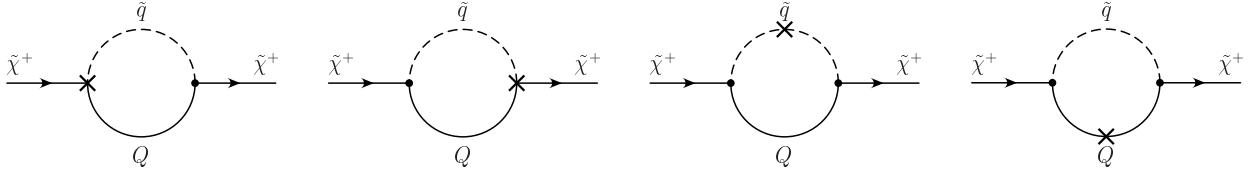}}}}
\end{picture}
\end{center}
\caption{\it  Chargino two-loop self-energy diagrams with counter-term insertions}\label{ChCT}
\end{figure}

Note that in contrast to the neutralino calculation there are now
quark isospin-partners denoted by $(q,Q)$ in the loop which give rise to
different tensor reduction formulae.
In order to get a pure $\alpha\alpha_S$ correction from (Fig.~\ref{Ch2LSquark})  it is necessary to shorten the
4-squark coupling to its QCD part.
Diagrams with one-loop counter-term insertions (Fig.~\ref{ChCT}) involve $\mathcal O (\alpha_S)$ mass counter-terms for quarks and squarks as well as
coupling constant counter-terms stemming from the Yukawa part of the chargino-quark-squark couplings and counter-terms to the squark mixing matrix,
see e.g. \cite{Bednyakov:2002sf}.

For the evaluation of the amplitudes we adopt the strategy of \cite{Schofbeck:2006gs}, that is, we use semi-automatic tools
\cite{feynarts, feyncalc, tarcer} in an {\sc Mathematica} environment and therein auto-create {\sc Fortran} code.

The renormalization prescription we adopt is the familiar \drbar-scheme which regulates UV-divergencies dimensionally but
introduces an unphysical scalar field for any gauge field in the theory in order to restore the counting of degrees of freedom in supersymmetry.
These unphysical mass parameters can be absorbed in the sfermion mass parameters~\cite{Martin:2001vx} (the resulting scheme is called $\drbarp$)
and hence provide a consistency check for the calculation of the diagrams containing gluon lines.

In order to handle infrared divergencies in the individual diagrams we introduce an infrared regulating mass parameter and check
that the resulting contributions as well as the unphysical scalar mass due to the gluon field cancel out in the final result.
As the main focus of this work is on the numerical analysis we do not reproduce the resulting lengthy expressions here.

Owing to the fact that we split the contributions into self-energy and counter-term diagrams, we can quite easily check some of the generic formulae
obtained in \cite{Martin:2005ch} where we find agreement.

The numerical analysis was performed by implementing  {\sc Tsil} \cite{tsil} in this {\sc Fortran} program.
As in the case of neutralinos we used the usual 't Hooft Feynman $R_{\xi=1}$ gauge for the gluon field, except for the check of gauge independence.

\begin{figure}[t]
\begin{center}
\begin{picture}(125,25)(0,0)
    \put(-19,-8){\mbox{\resizebox{!}{2.95cm}{\includegraphics{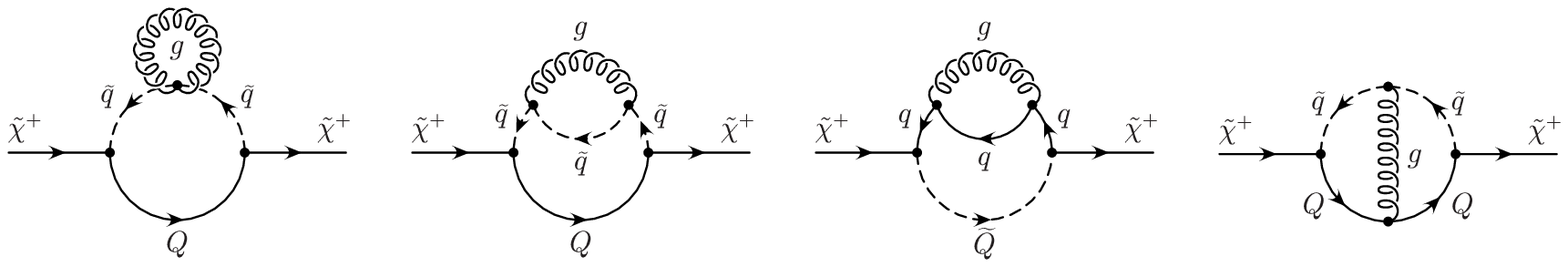}}}}
\end{picture}
\end{center}
\caption{\it Chargino two-loop selfenergy diagrams with inner gluon line}\label{Ch2LGluon}
\end{figure}

\begin{figure}[t]
\begin{center}
\begin{picture}(125,25)(0,0)
    \put(-4,-9){\mbox{\resizebox{!}{2.80cm}{\includegraphics{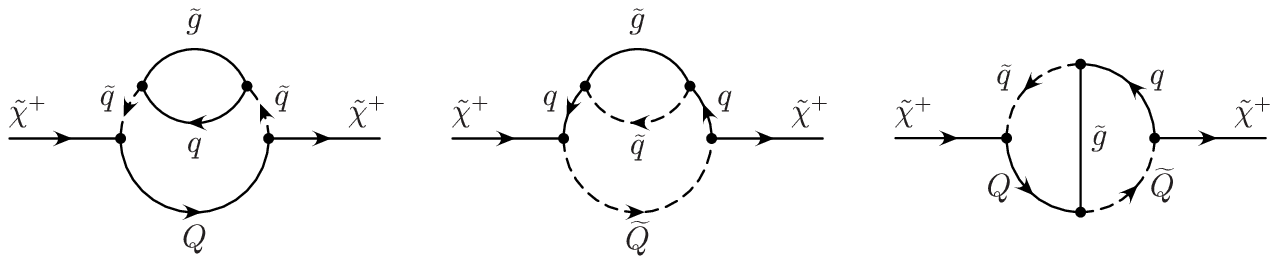}}}}
\end{picture}
\end{center}
\caption{\it  Chargino two-loop selfenergy diagrams with inner gluino line}\label{Ch2LGluino}
\end{figure}

\section{Numerics}

\begin{table}[t]
\begin{center}
\begin{tabular}{|c|c||c|c||c|}
    \hline
    \ \mbox{Particle} \ &
    \ \ \mbox{Mass}\ \  & \ \mbox{``LHC''}\ & \ \mbox{``ILC''}\
                        & \ \mbox{``LHC+ILC''}\ \\
    \hline\hline
    $\tilde{\chi}^0_1$    &  97.7 & 4.8  & 0.05 & 0.05 \\
    $\tilde{\chi}^0_2$    & 183.9 & 4.7  & 1.2  & 0.08 \\
    $\tilde{\chi}^\pm_1$  & 183.7 &      & 0.55 & 0.55 \\
    $\tilde{q}_R$         & 547.2 & 7-12 & -    & 5-11 \\
    $\tilde{q}_L$         & 564.7 & 8.7  & -    & 4.9  \\
    $\tilde{g}$           & 607.1 & 8.0  & -    & 6.5  \\ \hline
\end{tabular}
\end{center}
    \caption{accuracy from experiment at LHC, ILC and  $\textrm{LHC}\oplus\textrm{ILC}$ \cite{Weiglein:2004hn, spa}, masses in GeV}
    \label{acc}
\end{table}

Our reference scenario used for the numerical analysis is the benchmark point SPS1a' \cite{spa}. The SUSY parameters at $Q_0=1$~TeV are $\tan\beta = 10$,
$M_1 = 103.209$~GeV, $M_2 = 193.295$~GeV, $M_3 = 572.328$~GeV, $\mu = 401.62$~GeV, $A_t = -532.38$~GeV, $A_b = -938.91$~GeV,
$M_{\ti Q_3} = 470.91$~GeV, $M_{\ti U_3} = 385.32$~GeV and $M_{\ti D_3} = 501.37$~GeV , for further details see \cite{spa}. The tree-level chargino masses at this
point are $M_{\ti \chi_1^+} = 180.9$~GeV and $M_{\ti \chi_2^+} = 422.2$~GeV.

\begin{figure}[p]
 \begin{center}
\resizebox{!}{6cm}{\includegraphics{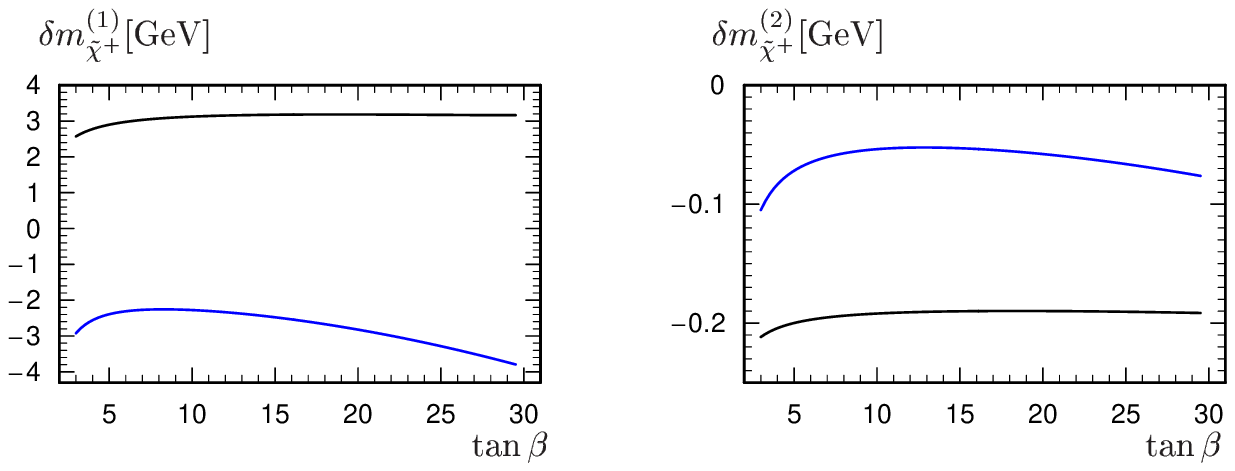}}
 \caption{Absolute chargino mass shifts as functions of $\tan\beta$. All other parameters are from SPS1a'.
 Left: One-loop mass shifts. Right: Two-Loop mass shifts.
 In all plots
 we use black for $\cha 1$ and blue for $\cha 2$.}\label{tanb}
 \end{center}
\end{figure}

Fig.~\ref{tanb} shows the chargino pole masses at SPS1a' as functions of $\tan\beta$. At the SPS1a' value $\tan\beta=10$ we find an absolute
two-loop correction $\delta m_{\cha 1} = 0.2\gev$ which is in the order of magnitude of the expected experimental uncertainty for this particle, see Table~\ref{acc}.
Therefore, the inclusion of these corrections is mandatory when extracting \drbar-parameters from experiment.

\begin{figure}[p]
\begin{center}
\hspace*{.3cm}
\resizebox{!}{6.1cm}{\includegraphics{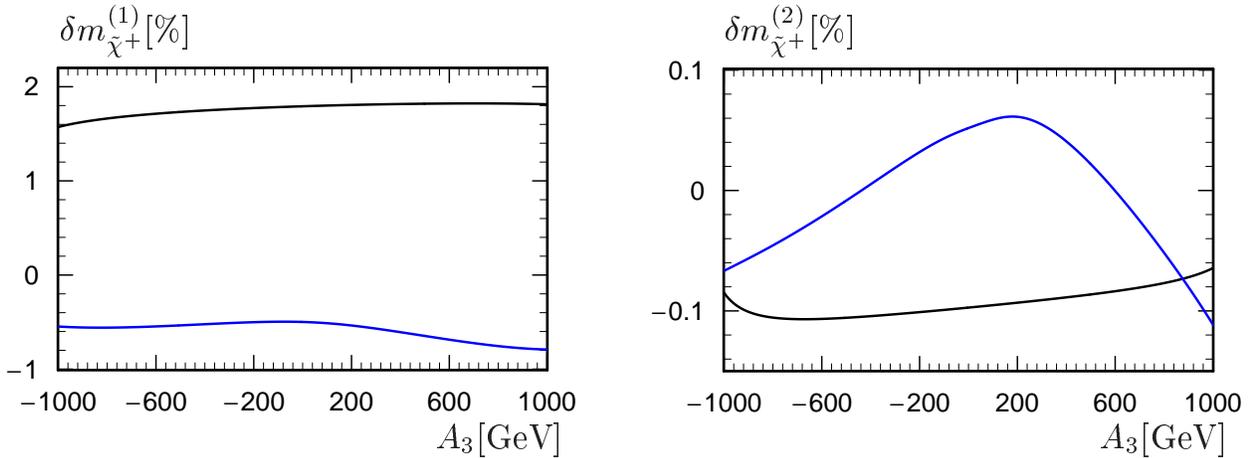}}
 \caption{\it Relative chargino mass shifts  $\delta m_{\cha 1}$ and $\delta m_{\cha 2}$ as functions
 of the trilinear breaking parameters $A_3 = A_t = A_b$. Left: One-loop mass shifts. Right: Two-Loop mass shifts.}\label{ATB}
 \end{center}
\end{figure}

For Fig.~\ref{ATB} we set the third generation trilinear breaking parameters equal, $A_3 = A_t = A_b$. This parameter effects the mixing
in the squark sector and therefore enters all the two-loop diagrams through the couplings and the sfermion masses.

\begin{figure}[p]
\begin{center}
\vspace*{-.5cm}
\resizebox{!}{5.8cm}{\includegraphics{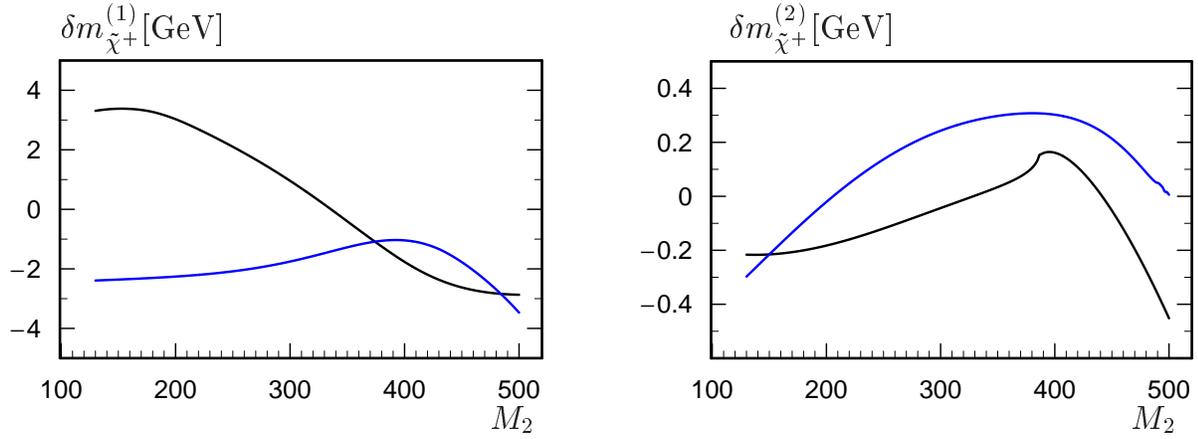}}
 \caption{Absolute chargino mass shifts  $\delta m_{\cha 1}$ and $\delta m_{\cha 2}$ as functions
 of the soft gaugino breaking mass $M_2$. In this plot we assume gauge unification, all other parameters are from SPS1a'.
  Left: One-loop mass shifts. Right: Two-Loop mass shifts.}\label{gaugeuni}
 \end{center}
\end{figure}

In Fig.~\ref{gaugeuni} we assume gauge unification, $M_1:M_2:M_3\simeq 1:2:6$. The plot is over $M_2$,
all other values are taken from SPS1a'.

\begin{figure}[p]
\begin{center}
\hspace*{.3cm}
\vspace*{-.5cm}\resizebox{!}{6.1cm}{\includegraphics{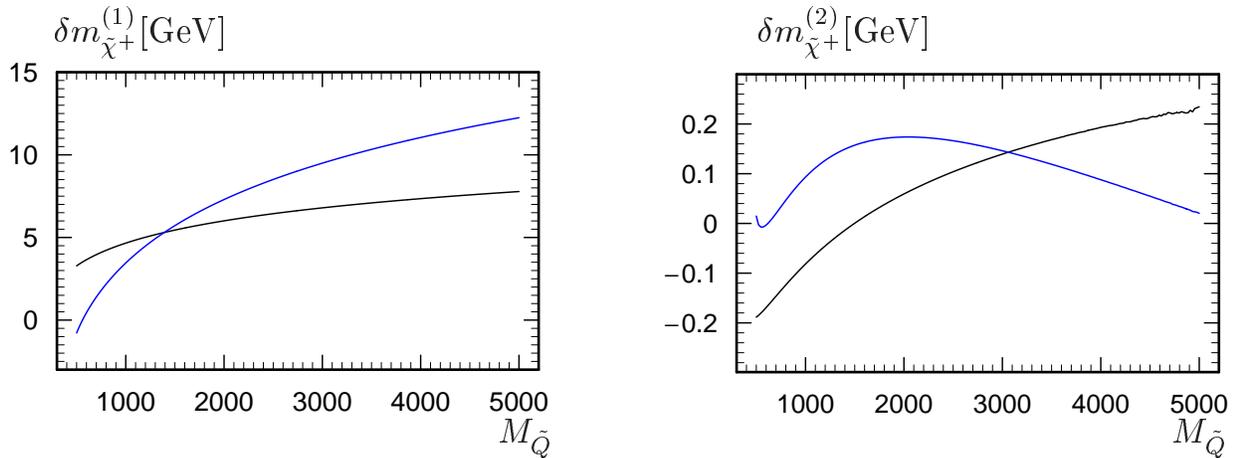}}
 \caption{Absolute chargino mass shifts as functions of $M_{\ti Q}$, see text. All other parameters are from SPS1a'.
 Left: One-loop mass shifts. Right: Two-Loop mass shifts.}\label{MSQ3}
 \end{center}
\end{figure}

In Fig.~\ref{MSQ3} we show the one- and two-loop chargino mass shifts as a function of the third generation soft SUSY breaking masses $M_{\ti Q}=M_{\ti Q_3}=M_{\ti U_3}=M_{\ti D_3}$.
Again, all other parameters are taken from SPS1a'.

\begin{figure}[p]
\begin{center}
\hspace*{-1cm}
\resizebox{!}{10cm}{\includegraphics{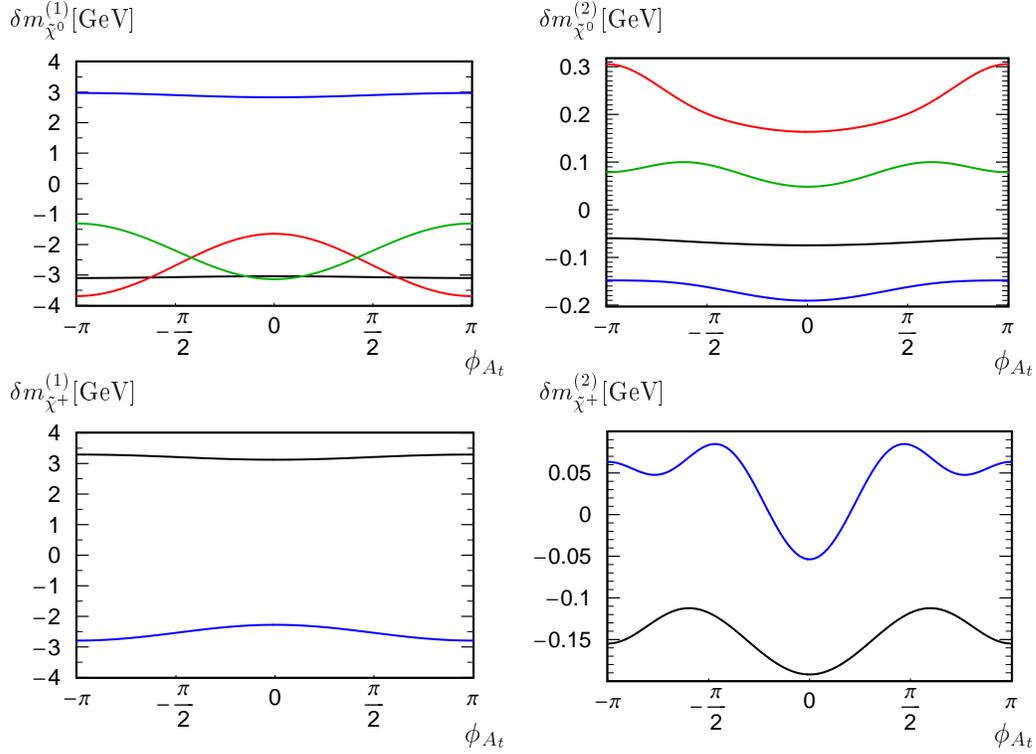}}
 \caption{Absolute neutralino and chargino mass shifts as functions of $\phi_{A_t}$, the complex phase of the soft trilinear breaking parameter $A_t$.
 First column: One-loop corrections. Second column: Two-loop corrections. The black line is $\ti \chi_1$, blue line is $\ti \chi_2$,
 red line is $\ti\chi^0_3$ and the green line is $\ti\chi^0_4$}\label{phAt}
\hspace{-1cm}
 \end{center}
\end{figure}

Finally, in Fig.~\ref{phAt} we investigate the dependence on $\phi_{A_t}$, the complex phase of the soft trilinear breaking parameter $A_t$.
Extending previous work \cite{Schofbeck:2006gs} we here include the neutralinos by taking $A_t$ complex.
In the first column there are the one-loop corrections to the chargino and the neutralino
 pole masses and in the second the respective two-loop corrections. It can be seen that the influence of the phase is quite substantial
at the two-loop level.

\begin{figure}[t]
\begin{center}
\resizebox{!}{6cm}{\includegraphics{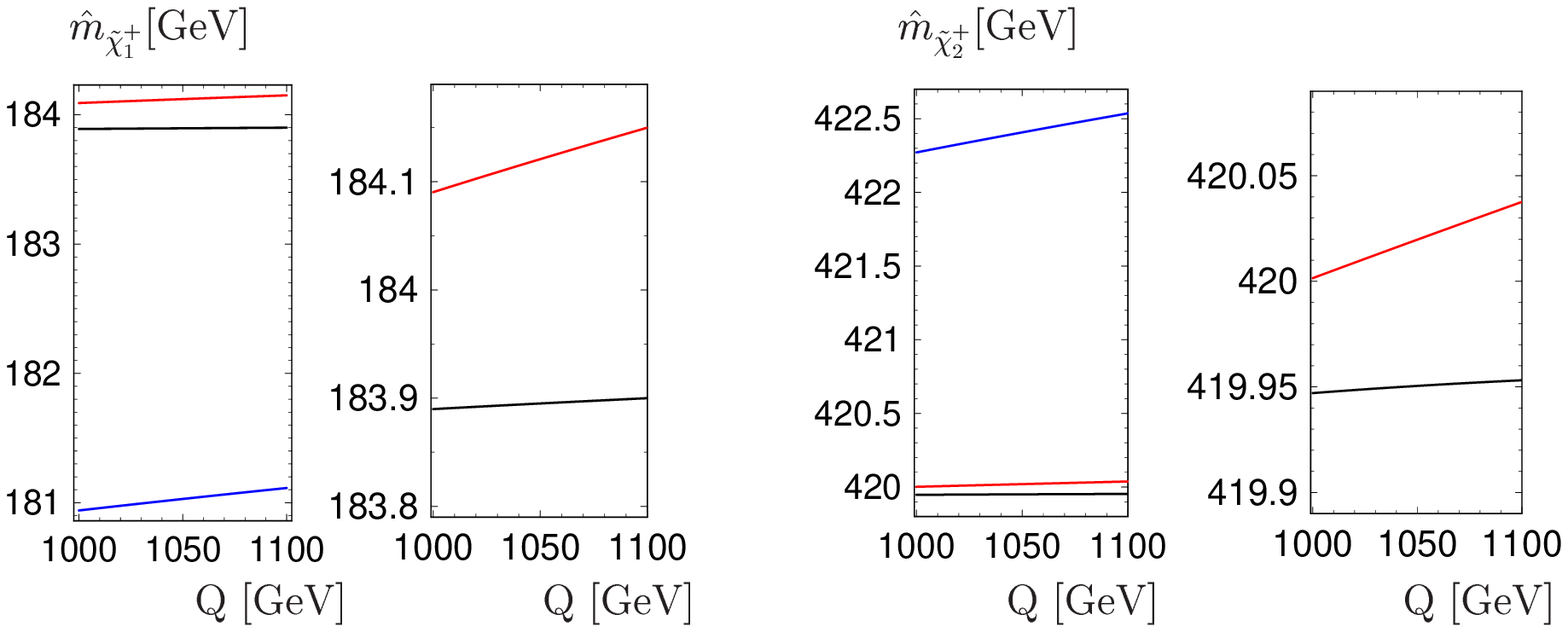}}
\hspace{-1cm}
 \caption{Scale dependence of chargino pole masses, see text. The blue line is the tree-level running $\drbar$-mass,
 red is the one-loop corrected mass and black the two-loop corrected mass}\label{scaledep}
 \end{center}
\hspace{-1cm}

\end{figure}

Fig.~\ref{scaledep} shows the decrease of the scale dependence when the loop-level of the corrections is increased.
The plots with one- (red) and two-loop (black) masses are zooms of the plots to their left where the running tree-level mass
is included. The scale dependence is reduced considerably when going from the one- to the two-loop level. The remaining scaling
comes from the uncancelled $\mathcal O(\alpha^2)$ RGEs.

\section{Conclusions}

We have calculated the chargino pole masses in the MSSM to order $\mathcal{O}(\alpha\alpha_S)$ and performed
a detailed numerical study. The typical size of the two-loop corrections is comparable to the expected eperimental accuracy at
future linear colliders and therefore needs to be taken into account when analyzing precision experiments. Our analytic expressions
agree with previous generic results \cite{Martin:2005ch} and have been checked thoroughly.
Extending previous work \cite{Schofbeck:2006gs} we also include complex parameters in the neutralino case.
Finally, we wrote a program \cite{Polxino} for this calculation with an
interface to the commonly used SUSY Les Houches accord \cite{Skands:2003cj}.

\appendix
\section{Polxino: Short description of the program}

Polxino \cite{Polxino} is a C-program for the evaluation of the pole masses of neutralinos and charginos in the MSSM. Up to now it contains
the full one-loop level and all SQCD effects at the two-loop level. It makes considerable use of the TSIL program \cite{tsil} and
uses some of the generic functions of \cite{gluinopole}. For
convenience we provide a SLHA-interface \cite{Skands:2003cj}.
We have carefully checked Polxino against the calculation described in this paper and in~\cite{Schofbeck:2006gs}.

\begin{table}[h]
\begin{tabular}{|l|l|}
\hline&\\
\texttt{\bf MSSM\_DATA} member & description\\&\\
\hline\hline&\\
 \texttt{\bf gs, g1,  g2} &  gauge coupling constants\\&\\
 \texttt{\bf mue} &  $\mu$-parameter\\&\\
 \texttt{\bf TB} & $\tan\beta$\\&\\
 \texttt{\bf VEV} &  \parbox[h]{7cm}{the square root of the sum\\ of squared Higgs VEVs}\\&\\
 \texttt{\bf MA02} & squared mass of $A_0$\\&\\
 \texttt{\bf MQU[3], MQD[3], MLE[3] } & quark and lepton-masses\\&\\
 \texttt{\bf M1, M2, M3 } & gaugino masses\\&\\
 \parbox[h]{7cm}{\texttt{\bf MSL[3], MSE[3], MSQ[3],\\MSU[3], MSD[3]}} & bilinear soft susy breaking parameters\\&\\
 \texttt{\bf Af[4][3]} & trilinear soft susy breaking parameters\\&\\
 \hline
\end{tabular}\caption{basic parameters of the struct \texttt{MSSM\_DATA}}
\label{basp}
\end{table}

The source code contains the main file {\bf main.c} which itself contains the code for a generic case, the evaluation of aforementioned pole masses
at the SPS1a' benchmark point. Since the program currently uses \texttt{\bf long double} (\texttt{\bf POL\_REAL}) and \texttt{\bf long double complex}
(\texttt{\bf POL\_COMPLEX}) it is recommended that TSIL is compiled with the compiler flag
\\[.3cm]\hspace*{2.5cm}\texttt{\bf -DTSIL\_SIZE\_LONG}\\[.3cm]
The central data struct in Polxino is
\\[.3cm]\hspace*{2.5cm}\texttt{\bf MSSM\_DATA}\\[.3cm]
which contains the whole set of MSSM parameters.
The subroutine loadSPC fills this struct from an SLHA-file and returns 1 if successful.
\\[.3cm]\hspace*{2.5cm}\texttt{\bf loadSPC("SPheno.spc",$\langle$MSSM\_DATA*$\rangle$);}\\[.3cm]
The file \texttt{\bf SPheno.spc} used by the main program was created by SPheno \cite{Porod:2003um} and comes with Polxino.
If the content of the struct \texttt{\bf MSSM\_DATA} is to be changed after the file was loaded it must be kept in mind that
some of the information stored in the struct is derived from a smaller set of basic elements.
The basic parameters which may be subject to change are given in Table~\ref{basp} with their respective index ranges in C/C++-notation.
The first index of \texttt{\bf Af} labels the
type of the trilinear breaking parameter according to {\sc FeynArts}. \texttt{\bf Af[0]} denotes the neutrino case and should therefore be kept zero.
\texttt{\bf Af[1]}, \texttt{\bf Af[2]} and \texttt{\bf Af[3]} are the leptonic, up- and down-type set of parameters. All other indices of
the form \texttt{\bf[3]} label the generation.

Calling
\\[.3cm]\hspace*{2.5cm}\texttt{\bf mssm\_digest($\langle$MSSM\_DATA*$\rangle$);}\\[.3cm]
recalculates all the derived parameters in the struct from the basic ones and re-diagonalizes the mixing systems. Thus, it is
fairly simple to loop over a parameter that enters in many different places, e.g. the $\mu$-parameter. The math used for this
re-diagonalization was extracted from some Fortran subroutines of {\sc LoopTools} \cite{Hahn:1998yk} and rewritten in the C-language.\\
The command
\\[.3cm]\hspace*{2.5cm}\texttt{\bf mssm\_print($\langle$MSSM\_DATA$\rangle$);}\\[.3cm]
prints out all the information stored in the struct. Finally,
\\[.3cm]\hspace*{2.5cm}\texttt{\bf getneupole($\langle$int$\rangle$ i, $\langle$MSSM\_DATA$\rangle$,}
\\\hspace*{4.95cm}\texttt{\bf $\langle$POL\_REAL*$\rangle$ Mpole1, $\langle$POL\_REAL*$\rangle$ Gamma1, }
\\\hspace*{4.95cm}\texttt{\bf $\langle$POL\_REAL*$\rangle$ Mpole2, $\langle$POL\_REAL*$\rangle$ Gamma2);}
\\[.3cm]\hspace*{2.5cm}\texttt{\bf getchpole($\langle$int$\rangle$ i, $\langle$MSSM\_DATA$\rangle$,}
\\\hspace*{4.65cm}\texttt{\bf $\langle$POL\_REAL*$\rangle$ Mpole1, $\langle$POL\_REAL*$\rangle$ Gamma1, }
\\\hspace*{4.65cm}\texttt{\bf $\langle$POL\_REAL*$\rangle$ Mpole2, $\langle$POL\_REAL*$\rangle$ Gamma2);}\\[.3cm]
calculate the one- and two-loop pole masses of chargino (neutralino) $i$ and store the one- respectivly two-loop results in the
variables \texttt{\bf Mpole1} resp.\texttt{\bf Mpole2}. The tree-level and one-loop widths are stored in \texttt{\bf Gamma1} and \texttt{\bf Gamma2},
respectively.

\clearpage
{\bf Acknowledgements}\\
The authors would like to thank W.~Majerotto, K.~Kova\v{r}\'{\i}k and
C.~Weber for discussion and many useful comments throughout the last year.
They especially thank W. Majerotto for his help in finalizing this work.
The authors acknowledge support from EU under the MRTN-CT-2006-035505
network programme. This work is supported by the "Fonds zur F\"orderung
der wissenschaftlichen Forschung" of Austria, project No. P18959-N16.

\end{document}